\RequirePackage{fix-cm}
\documentclass[twocolumn]{svjour3}          
\smartqed  
\usepackage[dvips]{graphicx}
\graphicspath{"D:\tex\"}

\usepackage{amsmath}
\usepackage{amsfonts}
\usepackage{amssymb}
\usepackage{cite}
\usepackage[numbers,sort&compress]{natbib}
\begin{document}

\title{Influence of constant electric field on circular photogalvanic effect in material with Rashba Hamiltonian
}


\author{V.I. Konchenkov,
        S.V. Kryuchkov,
        D.V. Zav'yalov
}


\institute{V.I. Konchenkov, S.V. Kryuchkov, D.V. Zav'yalov \at
		     Volgograd State Technical University\\ 
              400005, Lenina av., 28, Volgograd, Russia \\
              Tel.: +7-8442-248069\\
           \and
           S.V. Kryuchkov\at
           Physical laboratory of low-dimensional systems,\\
              Volgograd State Socio Pedagogical University \at
              400005, Lenina av., 27, Volgograd, Russia \\
              Tel.: +7-8442-949465\\
              \email{sed@fizmat.vspu.ru}
}

\date{Received: date / Accepted: date}

\maketitle

\begin{abstract}
An appearance of a direct current perpendicularly to a constant component of an electric field in material with Rashba Hamiltonian under the influence of an elliptically polarized wave is investigated in one-subband approximation. On its physical nature this effect is close to a circular photogalvanic effect (CPGE) on intraband transitions. The effect is studied on the base of two approaches: investigations of Boltzmann kinetic equation in a constant collision frequency approximation and semiclassical Monte Carlo simulations, which immediately takes into account microscopic processes of charge carriers scattering on optical and acoustical phonons. Monte Carlo modelling allows us to determine a mean relaxation time and its dependencies on electric field strengths and on the energy of optical phonons. On the base of these estimations a possibility of using a constant relaxation time approximation is justified. It is confirmed that the main contribution to the effects of transverse rectification is made by inelastic scattering of electrons on optical phonons. A comparison of results of Monte Carlo simulations and calculations on the base of a constant collision frequency approximation is presented.
\keywords{Rashba Hamiltonian \and non-additive energy spectrum \and circular photogalvanic effect \and Monte Carlo simulations}
\end{abstract}

\section{Introduction}
\label{intro}
A progress in telecommunications, using of new methods in data transfer, creation of medical equipment are in need of the development of microelectronic devices for generation, commutation and detection of different kinds of signals. The most actual problem is an engineering of equipments working in terahertz and far infrared range of electromagnetic waves (see, for example, \cite{1Andronov, 2UFNterahertz}). One way to determine the characteristics of high frequency electromagnetic waves (amplitude, frequency, phase) is a study of the direct current response which can occur under certain conditions in solid-state structures with non-linear properties (for example, \cite{{3Mensah, 4Shorokhov, 5Marchuk, 6Tulkina, 7Karch, 8GSLKontchenkov, 9KukharPhysE, 10KukharOptSpect, 11KukharFTT}}). Among the similar phenomena it should be noted a circular photogalvanic effect (CPGE) (see reviews \cite{12IvchenkoUFN, 13GanichevCondMat}), which was investigated in surface layers of bulk materials \cite{14GusevPJETP}, in thin films \cite{15Zhang}, quantum wells \cite{16Yu, 17Ganichev}, in graphene \cite{7Karch}. Without going into details, this effect is consisted in the appearance of a direct current under the influence of elliptically polarized electromagnetic wave with oblique incidence to the surface of the sample. The current flows perpendicularly to the projection of wave vector on the plane of the sample. By the magnitude of direct current one can draw a conclusion about properties of radiation as well as properties of a considered structure \cite{18Giglberger}. Effect is directly associated with the presence of spin of charge carriers and it was mostly explored in materials with Hamiltonian containing the term which is linear by a quasi-momentum (so-called Rashba \cite{19Rashba} or Dresselhaus \cite{20Dresselhaus} terms). Such Hamiltonians take in consideration a spin-orbital interaction immediately \cite{21Ganichev, 22Ganichev, 23Ivchenko, 24Tarasenko}. Most of the works on CPGE are devoted to research in visible and near infrared range of electromagnetic waves, where inter-band \cite{21Ganichev, 22Ganichev} or inter-subband \cite{17Ganichev, 23Ivchenko} optical transition are essential for kinetic properties of considered structures. In range of low frequencies of radiation, where only indirect intraband transitions without spin-flip are possible, the effect is not appear \cite{13GanichevCondMat}. It should be noted that even though photon energy is not enough for realization of direct transitions, there are possible transition processes through virtual intermediate state \cite{24Tarasenko}, which is placed in another band, and CPGE is feasible, but these transitions are second-order processes and the direct current magnitude in this case is noticeably less.

CPGE is related to the adsorption of circularly polarized electromagnetic wave which is resulting in optical orientation of spin because of angular momentum transfer from photons to electrons \cite{13GanichevCondMat}. A preferential direction appears in initially homogeneous electron momentum distribution which leads to a direct current arising. At oblique incidence of wave this preferential direction is defined by the projection of wave vector on the plane of the sample. At normal incidence of elliptically polarized wave a direct current is usually associated not so with symmetry breaking in distribution of charge carriers momenta as with the inhomogeneity of the sample \cite{7Karch}.

\begin{figure}
\includegraphics[scale=0.40]{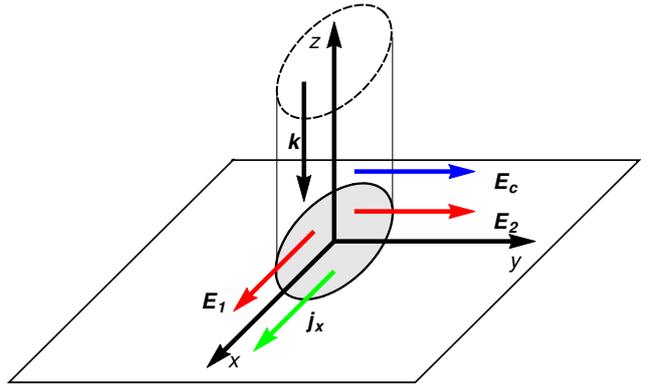}
\caption{Problem geometry}
\label{fig:1}       
\end{figure}

A preferential direction in material can be made by another way so, for example, applying a constant external electric field. In this situation a direct current generation is possible in direction perpendicular to the constant electric field under the influence of elliptically polarized electromagnetic wave, which is incident normally to the surface of the sample, when the frequency of wave is small compared with the band gap width. Herewith a main contribution to direct current component should be made by indirect transitions between the states within one band, that are the first-order processes, and the effect should not depend on spin. In papers \cite{5Marchuk, 6Tulkina, 8GSLKontchenkov} the effect of a direct current appearance in graphene and superlattice on the base of graphene in perpendicular to a constant field direction is investigated, when the elliptically polarized electromagnetic wave is incident normally to the surface of the sample (the problem geometry is shown in \ref{fig:1}). The effect arising relates to a material spectrum non-additivity \cite{5Marchuk, 8GSLKontchenkov}. In \cite{6Tulkina} on the base of Monte Carlo modelling it was shown that the inelastic scattering of charge carriers on optical phonons makes a main contribution in this effect. In present paper it is made an attempt to investigate on the base of semiclassical Monte Carlo simulations the effect of generation of a direct transverse current under a simultaneous influence of elliptically polarized wave and constant electric field in material described by Rashba Hamiltonian. Also we consider this problem in the approximation of a constant collision frequency and define the possibility of application of this approach.

\section{Problem statement}
\label{sec:1}
Let us to examine a sample which surface is irradiated by normally incident elliptically polarized electromagnetic wave with components of $\mathbf{E}_1$ and $\mathbf{E}_2$. A constant electric field $\mathbf{E}_c$ is applied along the surface of the sample (Fig. 1). The result electric field vector takes a form:
\begin{equation}
   \label{eq:eq1}
   \mathbf{E} = \{E_{10}\cos{\omega t},~E_c+E_{20}\cos{(\omega t + \varphi)}\}.
\end{equation}
We shall consider a two-dimensional material, described by the Rashba Hamiltonian:
\begin{equation}
   \label{eq:eq01}
   \hat{H}_R=\dfrac{\hat{p}^2}{2m}+\dfrac{\alpha}{\hbar}\left(\sigma_x \hat{p}_y-\sigma_y \hat{p}_x\right),
\end{equation}
where $\sigma_x,\ \sigma_y$ are the Pauli matrices:
\begin{equation}
   \label{eq:eq02}
  \sigma_x=\begin{pmatrix} 0 & \ 1 \\ 1 & \ 0 \end{pmatrix},\ \sigma_y=\begin{pmatrix} 0 & \ -i \\ i & \ \ 0 \end{pmatrix}. \nonumber
\end{equation}
Energy spectrum can be presented of:
\begin{equation}
   \label{eq:eq2}
   \varepsilon(\mathbf{p}) = \dfrac {\mathbf{p}^2}{2m}\pm\dfrac {\alpha}{\hbar}|\mathbf{p}|.
\end{equation}
Here $\alpha \sim 10^{-2} \mathrm{eV}\cdot\mathrm{nm}$ -- is a Rashba parameter \cite{18Giglberger, 25Cartoixa, 26Maiti, 27Nitta},
$ \mathbf{p} =\{ p_x, p_y \} $ -- is a quasi-momentum vector of electron. The expression (\ref{eq:eq2}) describes two subbands, in which a conductivity band splits after removal of the spin degeneracy \cite{13GanichevCondMat}. In present work we shall investigate transitions within one subband only, so we, to be more specific, shall choose in (\ref{eq:eq2}) plus sign. Successive accounting of the influence of interband or intersubband transitions on the effect will be a subject of a separate study.

Let us to find a direct component of the current density along the $Ox$  axis:
\begin{equation}
   \label{eq:eq3}
   j_x =  {q_{\mathrm{0}}n\langle v_x \rangle} ,
\end{equation}
where $v_x = \dfrac{p_x}{m} + \dfrac{\alpha}{\hbar} \dfrac{p_x}{\sqrt{p_x^2+p_y^2}}$
-- is a velocity projection, $n$ -- is a surface concentration of charge carriers ($n = N / L^2$, $L^2$ - surface area of crystal, $N$ - number of charge carriers), the angle brackets are denoted an averaging on ensemble and time. Strictly speaking, for carrying out averaging one has to know an electron distribution function, which requires the solution of kinetic equation. This approach we use in Sect. 3 where Boltzmann kinetic equation in a constant collision frequency approximation is solved. Because of mathematical complexity of problem the expression of direct current density is obtained in assumption of relative smallness of linear on quasi-momentum term in expression of energy spectrum in compare with quadratic term. This solution gives an opportunity to qualitative explain some features of the direct current density dependence on components of the electric field. But the approach of constant relaxation time doesn't take in account the microscopic processes of charge carriers scattering, so for definition of relative contribution of different types of scatterers on the direct current density we use Monte Carlo modelling. Method Monte Carlo allows us to perform an averaging of the velocity without the knowledge of an explicit form of the distribution function. Note, that used below Monte Carlo method is essentially a method of solving of Boltzmann kinetic equation (mathematically rigorous justification of this approach is presented, for example, in \cite{27Nedjalkov}), that is why all restrictions, related to the application possibility of the kinetic equation (transition probability assumed to be independent on applied fields, predominantly accounting pair collisions, etc.) concern to used approach too. The use of the Monte Carlo simulations for solving of physical kinetics problems (see, for example, \cite{28Grasser, 29Ting, 30Wang}) is a productive technique. Using a fairly simple and easy to parallelized code, one can get a sufficiently accurate and reasonable physical results. Sect. {\ref{sec:3}} is devoted to a description of the modelling method. Sect. {\ref{sec:4}} contains a calculation of charge carriers scattering probability on acoustical and optical phonons in material with Rashba Hamiltonian. Sect. {\ref{sec:5}} is devoted to a description of the modelling of charge carriers distribution on the quasi-momentum absolute value and polar angle in two-dimensional quasi-momentum space, which is used in Monte Carlo simulations. In Sect. {\ref{sec:6}} the results of Monte Carlo simulations are presented. Also in this section there is a comparison of Monte-Carlo simulation results and results of the investigation of the problem on the base of Boltzmann kinetic equation in approximation in constant collision frequency.

\section{Approach of a constant collision frequency}
\label{sec:2}
We shall investigate Boltzmann kinetic equation in approximation of a constant collision frequency:
\begin{equation}
   \label{eq:eq4}
   \dfrac{\partial f(\mathbf{p}, t)}{\partial t}+\mathbf{E} \dfrac{\partial f(\mathbf{p}, t)}{\partial \mathbf{p}} = - \nu\left(f(\mathbf{p},t) - f_0(\mathbf{p})\right).
\end{equation}
Here ${f_0}(\mathbf{p})$ is an equilibrium distribution function, $\nu$ is a mean collision frequency. The solution of (\ref{eq:eq4}) is a non-equilibrium distribution function. It can be found by a method of characteristics and takes a form
\begin{equation}
   \label{eq:eq5}
   f(\mathbf{p},t)=\nu \int\limits_{-\infty}^{t} dt'\exp(-\nu(t-t'))f_{0}(\mathbf{p'}(t';\mathbf{p},t)).
\end{equation}
Here $\mathbf{p'}(t;\mathbf{p},t)$ is a solution of semi-classical equations of motion
\begin{align}
   \label{eq:eq6}
   &\dfrac {dp'_x}{dt}=q_0E_{10} \cos(\omega t), \nonumber\\
   &\dfrac {dp'_y}{dt}=q_0E_c + q_0 E_{20} \cos(\omega t + \varphi).
\end{align}
The solution of (\ref{eq:eq6}) is
\begin{align}
   \label{eq:eq7}
   &p'_x = p_x - A_{0x}, \nonumber\\
   &p'_y = p_y - A_{0y},    
\end{align}
where we introduce denotations 
\begin{align}
   \label{eq:eq8}
   &A_{0x}=-\dfrac{q_0E_{10}}{\omega}(\sin\omega t'- \sin\omega t), \nonumber\\
   &A_{0y}=-q_0E_c(t'-t)- \nonumber \\
   &-\dfrac{q_0E_{20}}{\omega}(\sin(\omega t'+\varphi)- \sin(\omega t+\varphi)).    
\end{align}
We shall investigate non-degenerated electronic gas with equilibrium distribution function
\begin{equation}
   \label{eq:eq9}
   f_0(\mathbf{p})=A_{\mathrm{norm}}\exp\left(-\dfrac{\varepsilon(p_x,p_y)}{kT}\right).
\end{equation}
Here $T$ is absolute temperature, $A_{\mathrm{norm}}$ is a normalization constant,
\begin{equation}
   \label{eq:eq10}
   A_{\mathrm{norm}} = \dfrac{(2\pi\hbar)^2}{2\pi mkT}\cdot\dfrac{1}{\left(1-X\sqrt{\pi}e^{X^2}\left(1-\mathrm{erf}(X)\right)\right)},
\end{equation}
$\mathrm{erf}(x)$ is an error function, $X=\dfrac{\alpha m}{\hbar \sqrt{2mkT}}$ is a dimensionless parameter responsible for the spin-orbital interaction. So the current density along $Ox$ axis can be found by expression
\begin{align}
   \label{eq:eq11}
   &j_x = \dfrac{q_0 n}{(2\pi\hbar)^2} \int\limits_{-\infty}^{\infty} dp_x dp_y v_x(p_x,p_y)\cdot \nonumber\\
   &\cdot \int\limits_{-\infty}^{t}dt' \exp(\nu(t'-t))f_{\mathrm{0}}(p_x-A_{\mathrm{0}x},p_y-A_{\mathrm{0}y})).
\end{align}
Let us to carry out a substitution $p_x-A_{0x}\rightarrow p_x$, $p_y-A_{0y}\rightarrow p_y$. So the expression (\ref{eq:eq11}) is rearranged in the form
\begin{align}
   \label{eq:eq12}
   &j_x = \dfrac{q_0 n A_{\mathrm{norm}}}{(2\pi\hbar)^2} \int\limits_{-\infty}^{\infty} dp_x dp_y v_x(p_x+A_{\mathrm{0}x},{p_y+A_{\mathrm{0}y}})\cdot\nonumber\\
   &\cdot \int\limits_{-\infty}^{t}dt'\exp(\nu(t'-t))\exp\left(-\dfrac{\varepsilon(p_x,p_y)}{kT}\right).
\end{align}
In dimensionless denotations $(\mathbf{p}\rightarrow\mathbf{p}/\sqrt{2mkT},\ t\rightarrow\omega t)$ the current density is
\begin{align}
   \label{eq:eq13}
   &j_x = \dfrac{q_0 n \eta \sqrt{2mkT}}{\pi m}\cdot\dfrac{1}{\left(1-X\sqrt{\pi}e^{X^2}\left(1-\mathrm{erf}(X)\right)\right)}\cdot \nonumber\\
   &\int\limits_{-\infty}^{0} dt' \exp(\eta t')\int\limits_{-\infty}^{\infty}dp_x dp_y (p_x+A_x)\cdot \nonumber\\
   &\cdot\left(1+\dfrac{X}{\sqrt{(p_x+A_x)^2+(p_y+A_y)^2}}\right)\cdot \nonumber\\
   &\cdot\exp\left(-(p_x^2+p_y^2)-2X\sqrt{p_x^2+p_y^2}\right).
\end{align}
Here
\begin{align}
   \label{eq:eq14}
   &\eta=\dfrac{\nu}{\omega}, \ A_x=-E_x(\sin(t'+t)-\sin(t)), \nonumber\\
   \nonumber\\
   &A_y=-(E_{yc} t' + E_y(\sin(t'+t+\varphi)-\sin(t+\varphi))), \nonumber\\
   \nonumber\\
   &E_x = \dfrac{q_0E_{10}}{\omega\sqrt{2mkT}}, E_y = \dfrac{q_0E_{20}}{\omega\sqrt{2mkT}}, E_{yc} = \dfrac{q_0E_c}{\omega\sqrt{2mkT}}.
\end{align}
At a temperature $T\approx 70 K$, assuming $m\approx0.3\cdot 10^{-27}\mathrm{g}$, $\alpha\approx 0.6\cdot 10^{-2} \mathrm{eV}\cdot\mathrm{nm}$, the dimensionless Rashba parameter $X\approx0.12$, so in a first non-vanishing approximation on $X$ we can keep only a linear term in the expression of the direct current density. After carrying out the Taylor expansion on $X$ the expression under integral, passing into polar system of coordinates, remembering a subsequent averaging of the current density on time we give next expression for the current density:
\begin{align}
   \label{eq:eq15}
&j_x = j_0 \int\limits_{-\infty}^{0}dt'\exp(\eta t)\cdot \nonumber \\
&\cdot\int\limits_{0}^{\infty}\dfrac{A_x e^{-p^2}p^2dp}{\sqrt{p^2+{A_x}^2+{A_x}^2}\sqrt{{A_x}^2+{A_x}^2}}\cdot \nonumber\\
&\cdot\left(\int\limits_{0}^{\pi}d\phi\dfrac{\cos\phi}{\sqrt{1+b\cos\phi}}-\int\limits_{0}^{\pi}d\phi\dfrac{\cos\phi}{\sqrt{1-b\cos\phi}}\right),
\end{align}
\begin{equation}
   \label{eq:eq16}
j_0 = q_0 n \dfrac{\alpha}{\hbar}\dfrac{\eta}{\pi}, \ \ b = \dfrac{2p\sqrt{{A_x}^2+{A_y}^2}}{p^2+{A_x}^2+{A_y}^2}.
\end{equation}
According to Cauchy inequality $b\leq1$, so using \cite{31Prudnikov} (p.197, 1.5.20.8, 1.5.20.10) we receive:
\begin{align}
\label{eq:eq17}
   &\int\limits_{0}^{\pi}d\phi\dfrac{\cos\phi}{\sqrt{1+b\cos\phi}}-\int\limits_{0}^{\pi}d\phi\dfrac{\cos\phi}{\sqrt{1-b\cos\phi}}= \nonumber \\
   &=\dfrac{2}{b\sqrt{1+b}}\left[(1+b)\mathrm{E}\left(\sqrt{\dfrac{2b}{1+b}}\right)-2\mathrm{K}\left(\sqrt{\dfrac{2b}{1+b}}\right)\right. +\nonumber\\
   &+\left.(1-b)\mathrm{\Pi}\left(\dfrac{\pi}{2},\dfrac{2b}{1+b},\sqrt{\dfrac{2b}{1+ b}}\right)\right].
\end{align}
Here 
\begin{align}
\label{eq:eq18}
   &\mathrm{K}(k)=\int\limits_{0}^{\pi/2}\dfrac{1}{\sqrt{1-k^2\sin^2\theta}} d\theta, \nonumber \\
   &\mathrm{E}(k)=\int\limits_{0}^{\pi/2}\sqrt{1-k^2\sin^2\theta} d\theta \nonumber  
\end{align}
are complete elliptic integrals of the first and second kind, respectively, and
\begin{equation}
\label{eq:eq19}
   \mathrm{\Pi}(\varphi,n,k)=\int\limits_{0}^{\varphi}\dfrac{1}{(1-n\sin^2\theta)\sqrt{1-k^2\sin^2\theta}}d\theta \nonumber
\end{equation}
is an elliptic integral of the third kind.
After substitution ({\ref{eq:eq17}}) into ({\ref{eq:eq15}}) we get the expression of direct current density along $Ox$ axis which can be investigated further numerically:
\begin{align}
\label{eq:eq20}
   &{\left\langle j_x\right\rangle}_t = j_0\left\langle \int\limits_{-\infty}^{0}dt'e^{\eta t'}
   \cdot\right.\nonumber \\
   &\cdot\left.\int\limits_{0}^{\infty}dp\dfrac{e^{-p^2}p A_x \left(p^2+{A_x}^2+{A_y}^2\right)}{\left({A_x}^2+{A_y}^2\right)\left(p+\sqrt{{A_x}^2+{A_y}^2}\right)}\cdot\right.\nonumber \\
   &\cdot\left[
       (1+b)\ \text{E}\left(\sqrt{\dfrac{2b}{1+b}}\right)-2\text{K}\left(\sqrt{\dfrac{2b}{1+b}}\right)\right. \nonumber+ \\
   &{\left.+\left.(1-b)\mathrm{\Pi}\left(\dfrac{\pi}{2},\dfrac{2b}{1+b},\sqrt{\dfrac{2b}{1+b}}\right)\right]\right\rangle}_t.       
\end{align}

Angle brackets in a formula above denotes an averaging on a large in compare with a period of the elliptically polarized wave time interval. Because of time periodicity of the considered electric field we can perform averaging only by a period of wave. So for examine the dependence of the direct current component along $Ox$ axis we should numerically take three integrals (on absolute value of quasi-momentum $p$, time $t$ and $t'$).

\begin{figure}
\includegraphics[scale=0.45]{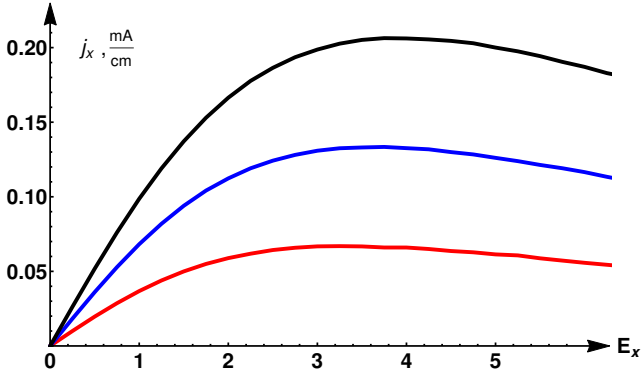}
\caption{Dependence of a transverse direct current density $j_x$ on a dimensionless component of electric field strength of elliptically polarized wave $E_x$. Red curve correspond to $E_y=1.0, E_{yc}=1.0$, blue curve -- to $E_y=1.5, E_{yc}=1.5$, black curve -- to $E_y=2.0, E_{yc}=2.0$.}
\label{fig:2}       
\end{figure}

\begin{figure}
\includegraphics[scale=0.45]{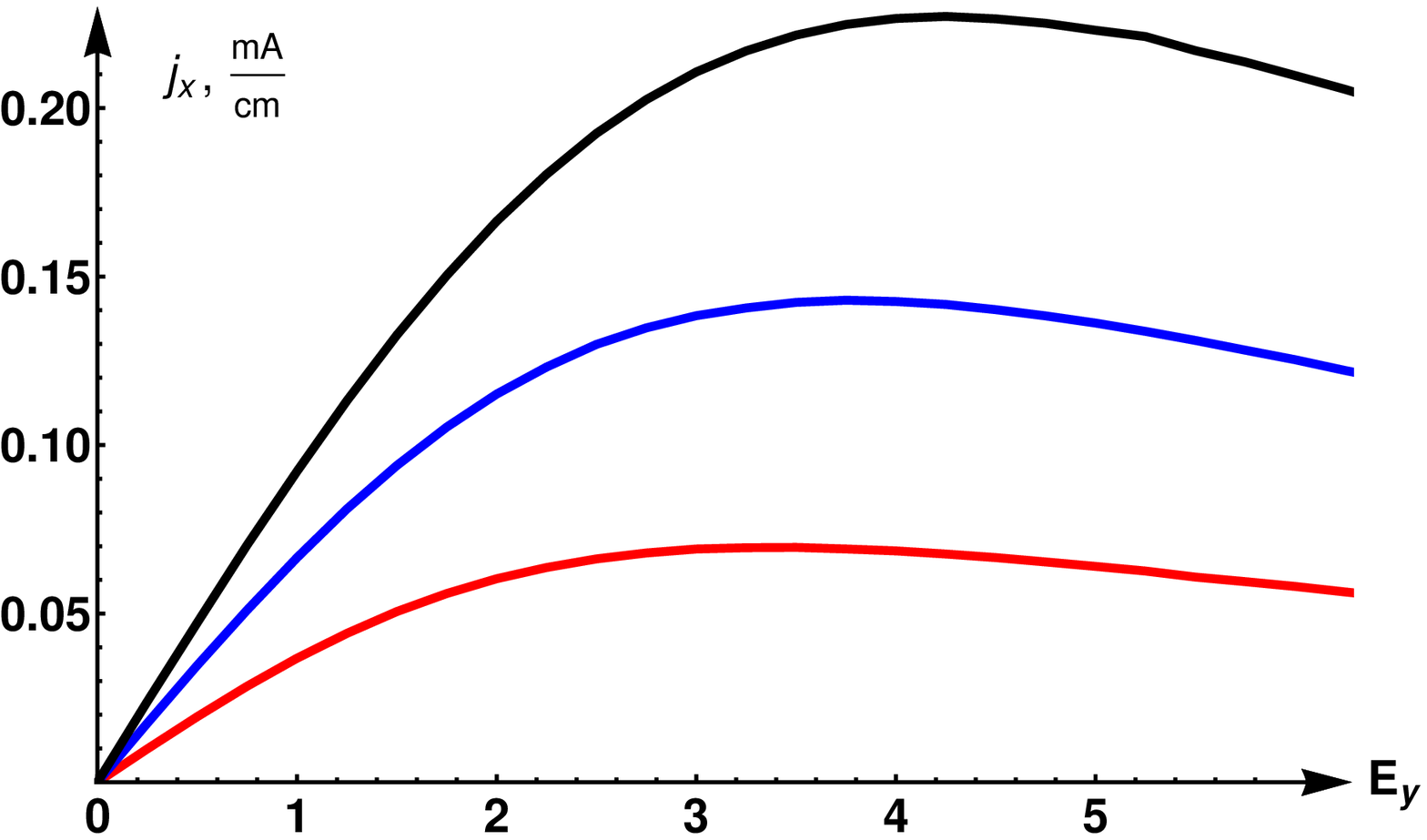}
\caption{Dependence of a transverse direct current density $j_x$ on a dimensionless component of electric field strength of elliptically polarized wave $E_y$. Red curve correspond to $E_x=1.0, E_{yc}=1.0$, blue curve -- to $E_x=1.5, E_{yc}=1.5$, black curve -- to $E_x=2.0, E_{yc}=2.0$.}
\label{fig:3}       
\end{figure}

\begin{figure}
\includegraphics[scale=0.45]{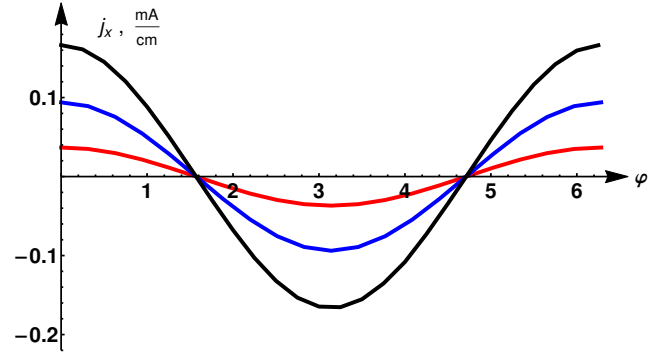}
\caption{Dependence of a transverse direct current density $j_x$ on the phase shift $\varphi$ between the components of elliptically polarized wave. Red curve correspond to $E_x=1.0, E_y=1.0, E_{yc}=1.0$, blue curve -- to $E_x=1.5, E_y=1.5, E_{yc}=1.5$, black curve -- to $E_x=2.0, E_y=2.0, E_{yc}=2.0$.}
\label{fig:4}       
\end{figure}

Figures {\ref{fig:2}} and {\ref{fig:3}} illustrate the dependence of the transverse direct current density $j_x$ on components of the electric field strength of elliptically polarized wave $E_x$ and $E_y$, respectively. Value $E_x=1$ ($E_y, E_{yc} = 1$) corresponds to approximately 10 SGSE unit at a temperature $T=70\mathrm{K}$, cyclic frequency of incident wave $\omega\simeq 10^{12} \mathrm{s}^{-1}$. Rashba parameter is assumed to be $\alpha = 0.8\cdot 10^{-2} \mathrm{eV}\cdot \mathrm{nm}$. In section \ref{sec:6} on the base of Monte Carlo simulation we find a mean collision frequency $\nu\simeq 10^{13} \mathrm{s}^{-1}$, so graphics on Figures {\ref{fig:2}} and {\ref{fig:3}} are plotted in assumptions $\eta=\nu/\omega=10$. Also, to be more specific, we assume $\varphi=0$. The dependencies of the direct current on a constant component of electric field $E_{yc}$ are similar ones presented on Figures {\ref{fig:2}} and {\ref{fig:3}}, and they are shown in Section {\ref{sec:6}}, where the results of the constant collision frequency approach and the Monte Carlo modelling are compared. In a low-amplitude area of electric field the direct current density is linear with respect to each component $E_x$, $E_y$, $E_{yc}$. This result is in according with the same obtained previously in similar problem geometry in graphene analytically \cite{5Marchuk} and by Monte Carlo simulation \cite{6Tulkina}. The effect is cubic by magnitudes of electric field strengths which is closely related to mutual influence of charge carriers motions in transverse each other directions in material with non-additive energy spectrum. A non-monotonic character of a direct current density dependence on the components of the electric field can be explained that the effect is result of a mutual interaction of three motions (under the influence of two alternative and one constant forces). So if one of the forces became much more than two another, the effect decreased. Another cause of decreasing of the direct current with rising of one component of the electric field is that in the states where an electron has small values of quasi-momentum, a term in the energy spectrum,  linear with respect to a quasi-momentum, prevails over the quadratic term, and a non-additivity of the energy spectrum is significant. With rising of a mean value of a quasi-momentum a relative contribution of the linear term is reduced, an influence of the energy spectrum non-additivity decreases and the direct component of a transverse current also diminishes.

The dependencies of the direct current density on a phase shift between the components of wave is very close to cosine (Fig. {\ref{fig:4}}). With growth of a collisional frequency the direct current component monotonically decreases. 

Consequently, on the base of examination of Boltzmann kinetic equation with the collision term in a form of constant relaxation frequency we calculate a direct current density (in perpendicular direction to a constant component of electric field strength) in material with Rashba Hamiltonian under the influence of elliptically polarized electromagnetic wave, which is incident normally to the surface of the sample. This approach allows us to get the dependence of direct current on components of electric field, collision frequency and phase shift between the components of elliptically polarized wave. It is established that essential contribution in effect appearance belongs to energy spectrum non-additivity. But the method of constant relaxation time can't immediately take in account microscopic mechanisms of charge carriers collisions on the lattice heterogeneity. Meanwhile, only a non-additivity can't lead to appearance a preferential direction in a sample. Direct current should arise as result of momentum and energy interchange between electron gas and lattice under the influence of electric field. Next sections are devoted to clarification of relative influence of different types of scattering on appearance of direct current on the base of semi-classical Monte Carlo simulations.  
 
\section{Description of modelling method}
\label{sec:3}
Let us to watch over an electron which moves in electric field and is capable to be scattered by lattice irregularities. We assume that during the intervals between collisions the particle motion obeys the laws of classical mechanics and is described by the equations of motion
\begin{equation}
   \label{eq:eq23}
   \dfrac {d\mathbf{p}}{dt} =  q_0\mathbf{E}(t).
\end{equation}
The time intervals between collisions and the final momentum acquired by the particle after the collision, we model by the random variable, the distribution of which will be determined by the probabilities of transition from one state to another as a result of scattering. The transition probabilities are calculated from quantum mechanical considerations.

Calculating the momentum component for single electron at each point of time (taking into consideration the motion in electric field and scattering), we calculate the electron velocity in each point of time. Repeating many times similar calculations for different initial conditions (one might say, for different particles), and performing an averaging across the set of initial conditions and with respect to a time, we obtain an expression for the mean velocity, and then, using (\ref{eq:eq3}), we calculate a direct current density.

As it was shown by simulation \cite{6Tulkina} and as it follows from physical reasons, an inelastic scattering should play a key role in causing of transverse rectification effects. An energy exchange between the lattice and electron gas leads to demonstration of the mutual dependence of charge carriers movements in directions perpendicular to each other, potentially inherent in the materials with non-additive energy spectrum. That is why during the farther consideration of effect of direct current appearance in material with Rashba Hamiltonian under the conditions of elliptically polarized electromagnetic wave we shall take in account two mechanisms: scattering of charge carriers on optical phonons (substantially inelastic) and scattering on acoustical phonons (which is close to elastic scattering). We shall need the expressions of total transition probability $W(\mathbf{p})$ from the state with momentum $\mathbf{p}$ to the state with any another allowable momentum value. The Sect.~\ref{sec:6}~ is devoted to calculation of these probabilities. This section describes the considerations, on the base of which the time of electron scattering on phonons is modelled and the scattering mechanism is chosen.

A mean free time can be found by the next considerations set forth, for example, in \cite{31Kireev}. Suppose, first, that the probability of an electron scattering in a time interval is proportional to the interval, and, second, the probability of collisions per unit time is independent of time. Let $w(dt)$ is a probability that a particle moving without collisions over a period of time $(t,\ t+dt)$. Consider the event $C$, consisting in the fact that electrons move without scattering in the interval $(t_0, t_0+t+dt)$. This event can be considered as a combination of two events: the event $A$, which consists in the fact that the electron moves without scattering over a period of time $(t_0,\ t_0+t)$, and the event $B$ consisting in the fact that the electron is moving without being scattered in time interval $(t_0+t,\ t_0+t+dt)$. Since we assume that the probability of the free motion of a particle depends on the duration of the period of time, but not on the initial (or final) point of time, the events $A$ and $B$ are independent, so
\begin{equation}
   \label{eq:eq24}
   w(t+dt)=w(t)\cdot w(dt).
\end{equation}
On the other hand,
\begin{equation}
   \label{eq:eq25}
   w(t+dt)=w(t)+\dfrac{dw}{dt}dt.
\end{equation}
The probability of free movement for the time $dt$  can be expressed by the scattering probability $W$ for the same time:
\begin{equation}
   \label{eq:eq26}
   w(dt) = 1 - Wdt.
\end{equation}
For $w(t)$ one obtains next equation:
\begin{equation}
   \label{eq:eq27}
   w(t) + \dfrac{dw}{dt}dt = w(t)(1-Wdt).
\end{equation}
The solution of this equation is a function
\begin{equation}
   \label{eq:eq28}
   w(t) = A\exp(-Wt),
\end{equation}
where the constant $A$ is determined by the normalization condition
\begin{equation}
   \label{eq:eq29}
   \int\limits_{0}^{\infty}w(t)dt  = 1.
\end{equation}
Thus, the probability that an electron will move without scattering within the time $(t_0,\ t_0+t)$
\begin{equation}
   \label{eq:eq30}
   w(t)= W\exp(-Wt).
\end{equation} 
There $W$ is a total probability of charge carriers scattering per unit time. This probability depends primarily on the electron quasi-momentum: $W=W(\mathbf{p})$. The quantity $Wt$ in the exponent, in fact, is the probability of charge carriers scattering in the time interval duration $t$. When the sample is placed in an electric field, an electron moves between collisions according to the equation of motion (\ref{eq:eq23}), so that the momentum (and the probability of scattering) begins to depend on the time. Generalization of (\ref{eq:eq30}) in this case is to replace
\begin{equation}
   \label{eq:eq31}
   \mathbf{p}\rightarrow \mathbf{p}(t)=\mathbf{p} - \dfrac{e}{c}\mathbf {A}(t)
\end{equation}
($\mathbf{A}(t)=-c\int\limits_{0}^{t}\mathbf{E}(t)dt$ is a vector-potential),
\begin{equation}
   \label{eq:eq32}
   Wt \rightarrow\int\limits_{0}^{t} W(\mathbf{p}(t'))dt',
\end{equation}
so the probability that an electron will not experience a collision during a period $t$ when it moves under the influence of the external field is
\begin{equation}
   \label{eq:eq33}
   w(t)=\int\limits_{0}^{t} W(\mathbf{p}(t'))\exp\left(-\int\limits_{0}^{t'} W(\mathbf{p}(t''))dt''\right)dt'.
\end{equation}
Expression (\ref{eq:eq33}) can be used for drawing the time between successive scatterings. We assign to $w(t)$ a value  $r$ uniformly distributed in the interval $(0,\ 1)$ . Then the time $t$ of motion of an electron under the influence of fields between two successive collisions can be determined by solving the equation
\begin{equation}
   \label{eq:eq34}
   r=\int\limits_{0}^{t} W(\mathbf{p}(t'))\exp\left(-\int\limits_{0}^{t'} W(\mathbf{p}(t''))dt''\right)dt'.
\end{equation}
Solving of (\ref{eq:eq34}) is a complex process. Therefore, to simplify the calculations, we use techniques developed in \cite{32Rees1, 33Rees2}. We introduce a fictitious self-scattering process, which does not change the momentum of a particle \cite{34Hockney}. Let the probability of transition between the states $\mathbf{p}$ and $\mathbf{p'}$ as a result of this process
\begin{equation}
   \label{eq:eq35}
   W_0(\mathbf{p}, \mathbf{p'})=W_{\mathrm{0}}(\mathbf{p})\delta(\mathbf{p}-\mathbf{p'}).
\end{equation}
Value $W_0(\mathbf{p})$ in (\ref{eq:eq35}) can be chosen arbitrarily, since the state of the particle as a result of self-scattering does not change. In (\ref{eq:eq34}) $W(\mathbf{p})$ is the probability of scattering due to all the processes taken into consideration. Then, with the introduction of self-scattering, (\ref{eq:eq34}) takes a form:
\begin{align}
   \label{eq:eq36}
   &r=\int\limits_{0}^{t} dt'(W(\mathbf{p}(t'))+W_{\mathrm{0}}(\mathbf{p}(t'))) \cdot \nonumber \\ 
   &\cdot\exp\left(-\int\limits_{0}^{t'}(W(\mathbf{p}(t''))+W_{\mathrm{0}}(\mathbf{p}(t'')))dt''\right).
\end{align}
Considering $W_0(\mathbf{p})=\mathrm{\Gamma} - W(\mathbf{p})$ where $\mathrm{\Gamma}$ is some constant, we obtain the condition for the determination of the collision moment in the form of
\begin{equation}
   \label{eq:eq37}
   r = 1- \exp(-\mathrm{\Gamma}t).
\end{equation}
The value $\mathrm{\Gamma}$ should be chosen in such a way that in the region of interest values of the momentum the value $W_0(\mathbf{p})$  was positive.

There is another approach to the calculation of the mean free time \cite{35Sobol'}. As before, we assume that the probability of scattering particles in the time interval $(t,\ t+dt)$  is proportional to the length of this interval:
\begin{equation}
   \label{eq:eq38}
   w_{\mathrm{scat}}(dt) = Wdt.
\end{equation}
Here $W$ is the probability of charge carriers scattering per unit time. Let the random variable $r$ is a time of free path. We introduce the distribution function of this value $F(t)$ -- the probability that the time of free path $\tau$  is less than $t$:
\begin{equation}
   \label{eq:eq39}
   F(t)=\mathrm{P}\{\tau < t \}.
\end{equation} 
Then the probability of the event $C$, consisting in the fact that the time of free path is in between $(t,~t+dt)$, will be equal   
\begin{equation}
   \label{eq:eq40}
   \mathrm{P}(C) = F(t+dt)-F(t).
\end{equation}
On the other hand, the event $C$ can be represented as a superposition of two events: $A$ -- "the particle is moving without scattering over a period of time $(0,~t)$" and $B$ -- "particle scatters in $(t,~t+dt)$". These events are independent. An independence of event $A$ on the event $B$ is unquestioned. An independence of event $B$ on $A$ is clear from the following. The fact that the scattering of particles is occurred in the interval $(t,~t+dt)$ (event $B$) by our assumption depends on the length of the interval $dt$  and on the state of the particles (primarily, on its quasi-momentum). Electron scattering probability per unit time $W$ depends on the quasi-momentum of particle, which in the case of external fields applying is a function of time, but there is no explicit dependence of the probability on the time measured from the previous collision, so there is no explicit dependence of the event $B$ on the initial time point of interval $(t,~t+dt)$, and, hence, there is no the dependence of the event $B$ on the event $A$. Therefore the probability of an event $C$ can be represented as a product of probabilities of the events $A$ and $B$:
\begin{equation}
   \label{eq:eq41}
   \mathrm{P}(C) = \mathrm{P}(A)\cdot \mathrm{P}(B).
\end{equation}
Obviously, the probability of event $A$ is a complementary quantity for $F(t)$:
\begin{equation}
   \label{eq:eq42}
   \mathrm{P}(A) = 1-F(t).
\end{equation}
The probability of event $B$ is defined by (\ref{eq:eq38}). Combining (\ref{eq:eq38}), (\ref{eq:eq40}), (\ref{eq:eq41}), (\ref{eq:eq42}), we obtain the following relationship:
\begin{equation}
   \label{eq:eq43}
   F(t+dt)-F(t)=(1-F(t))\cdot Wdt.
\end{equation}
Thus, the distribution function of the time of free path satisfy the differential equation
\begin{equation}
   \label{eq:eq44}
   \dfrac{dF}{dt}=W(1-F(t)).
\end{equation}
A solution of that equation in case of $W=W(\mathbf{p}(t))$ takes a form:
\begin{equation}
   \label{eq:eq45}
   F(t)=1-\exp\left(-\int \limits_{0}^{t}W(\mathbf{p}(t'))dt'\right).
\end{equation}
In modelling the expression (\ref{eq:eq45}) can be used to determine the time of free path, assuming $F(t)$ is equal to uniformly distributed on the interval $(0,~1)$ random variable $r$. Thus, the time of free path is determined by solving the equation
\begin{equation}
   \label{eq:eq46}
   r=1-\exp\left(-\int\limits_{0}^{t}W(\mathbf{p}(t'))dt'\right).
\end{equation}
The expression (\ref{eq:eq46}) is more preferable then the expressions (\ref{eq:eq34}), (\ref{eq:eq37}), since during the derivation of (\ref{eq:eq34}) we first assumed $W$ was independent on time, and then we generalized (\ref{eq:eq30}) (not so strict) for the case of an implicit dependency of the scattering probability $W(\mathbf{p}(t'))$ on time. The expression (\ref{eq:eq37}) takes into account the dependence of the time of free path on the scattering probability only indirectly. For practical calculations it is convenient to transform the expression (\ref{eq:eq45}):
\begin{equation}
   1-F(t)=\exp\left(-\int \limits_{0}^{t}W(\mathbf{p}(t'))dt'\right).\tag{42'}
\end{equation}
Left part in (42') is the probability of the fact that the time of free path is greater than $t$, and this probability one can simulate by a uniformly distributed random variable $r'$:
\begin{equation}
   \label{eq:eq48}
   r'=\exp\left(-\int\limits_{0}^{t}W(\mathbf{p}(t'))dt'\right).\tag{42''}
\end{equation}
In a right part of (42'') under the exponent integral is taken numerically (trapezoidal method):
\begin{equation}
   \label{eq:eq49}
   r=\exp\left(-\dfrac{\Delta t}{2}\sum \limits_{i=0}^{n-1} \left(W(\mathbf{p}(t_{i+1}))+W(\mathbf{p}(t_i))\right)\right),
\end{equation}
where $\Delta t$ is a time step (we have omitted the quote in the designation of uniformly distributed random variable). Since a right part of (\ref{eq:eq49}) is a monotonous decreasing function, we find a time of free path according to
\begin{equation}
   \label{eq:eq50}
   \tau=\Delta t\cdot n,
\end{equation}
where $n$ is minimal number of steps required to satisfy the condition
\begin{equation}
   \label{eq:eq51}
   r\ge \exp\left(-\dfrac{\Delta t}{2}\sum \limits_{i=0}^{n-1} \left(W(\mathbf{p}(t_{i+1}))+W(\mathbf{p}(t_i))\right)\right).
\end{equation}
Taking the logarithm of both sides of (\ref{eq:eq51}), one gets following condition
\begin{equation}
   \label{eq:eq52}
   -\ln r\le \dfrac{\Delta t}{2}\sum\limits_{i=0}^{n-1} \left(W(\mathbf{p}(t_{i+1}))+W(\mathbf{p}(t_i))\right),
\end{equation}
which is used in this paper to perform calculations.

Playing a random variable $r$ uniformly distributed in the interval $[0,~1]$, we find a moment of collision using (\ref{eq:eq52}). Then one should select the scattering mechanism. Thereto one more random variable $s$, uniformly distributed in the interval $[0,~1]$, is played out. In a case when only two mechanisms of scattering are supposed (scattering by acoustic and optical phonons), the choice of the scattering mechanism is based on the next obvious condition: one assumes that scattering was happened by optical phonon, if
\begin{equation}
   \label{eq:eq53}
   \dfrac {W_{opt}(\mathbf{p})}{W_{opt}(\mathbf{p})+W_{ac}(\mathbf{p})} \ge s,
\end{equation}
and one assumes that electron was scattered by acoustic vibrations of the lattice in a case if
\begin{equation}
   \label{eq:eq54}
   \dfrac {W_{opt}(\mathbf{p})}{W_{opt}(\mathbf{p})+W_{ac}(\mathbf{p})} < s.
\end{equation}
Defining the scattering mechanism, we calculate the momentum after the collision on the base of the laws of conservation of energy and momentum, further we estimate the velocity at any time during the motion of an electron in an external field and define a new collision moment.

\section{The probabilities of charge carriers scattering on acoustical and optical phonons}
\label{sec:4}
\subsection{Charge carriers scattering probability on long-wavelength nonpolar phonons in two-dimensional materials with linear on quasi-momentum term in energy spectrum}
\label{sec:4.1}
Let us to consider a probability of the electron transition from the state with a quasi-momentum $\mathbf{p}$ to the state $\mathbf{p'}$\cite{36Bonch}:
\begin{align}
   \label{eq:eq55}
   &W\left(\mathbf{p};\ \mathbf{p'}\right)= \nonumber \\
   &=\dfrac {2\pi}{\hbar}\sum\limits_{n'}\left|\left\langle \mathbf{p'},n'\left|H'\right|\mathbf{p}, n \right\rangle\right|^2\cdot\delta\left(\epsilon\left(\mathbf{p'}, n'\right)-\epsilon\left(\mathbf{p}, n\right)\right).
\end{align}
Here $H'$ - is the operator of electron-phonon interaction (it depends on type of phonons), matrix element $\left|\left\langle \mathbf{p'}, n'\left|H'\right|\mathbf{p}, n \right\rangle\right|$ is taken on wave function of a system "electron + phonons", $n$ is a number of phonons in a chosen state, $\epsilon$ is an energy of the system "electron + phonons" in this state. The operator of electron-phonon interaction takes a form \cite{36Bonch}
\begin{equation}
   \label{eq:eq56}
   H'=\sum\limits_{\mathbf{q}, s}\left\lbrace H'\left(\mathbf{q}, s\right)b(\mathbf{q}, s)e^{i\mathbf{qr}}+H'^{*}\left(\mathbf{q}, s\right)b^{*}(\mathbf{q}, s)e^{-i\mathbf{qr}}  \right\rbrace,
\end{equation}
where $b^{*}(\mathbf{q}, s)$ and $b(\mathbf{q}, s)$ are, respectively, birth and delete operators of phonon with a branch number $s$ and a quasi-momentum vector $\mathbf{q}$. Wave function $\left|\mathbf{p}, n\right\rangle$ relates to a system of non-interacting electron and phonon gases, so it can be presented in a form of the product
\begin{equation}
   \label{eq:eq57}
  \left|\mathbf{p}, n\right\rangle = \mathrm{\Psi}_{\mathbf{p}}\left(\mathbf{r}\right)\mathrm{\Phi}_{n},
\end{equation}
here $\mathrm{\Psi}_{\mathbf {p}}\left(\mathbf{r}\right)$ - is an electron wave function, $\mathrm{\Phi}_n$ - is a phonon wave function which is equal to a wave function of harmonic oscillator. So the expression of matrix element can be presented as
\begin{align}
   \label{eq:eq58}
  &\left\langle \mathbf{p}',n'\left|H'\right|\mathbf{p}, n \right\rangle = \nonumber \\
  &=\sum\limits_{\mathbf{q}, s} H'\left(\mathbf{q}, s\right)J_1\int\mathrm{\Phi}_{n'} b(\mathbf{q}, s)\mathrm{\Phi}_{n}\prod\limits_{\mathbf{q}'',s''}dx_{\mathbf{q}'',s''}+ \nonumber \\
 &+\sum\limits_{\mathbf{q}, s} H'^{*}\left(\mathbf{q}, s\right)J_2\int\mathrm{\Phi}_{n'} b^{*}(\mathbf{q}, s)\mathrm{\Phi}_{n}\prod\limits_{\mathbf{q}'',s''}dx_{\mathbf{q}'',s''}.
\end{align}
Here
\begin{equation}
   \label{eq:eq59}
  J_1 = \int d\mathbf{r}\mathrm{\Psi}^{*}_{\mathbf{p}'}e^{i\mathbf{qr}}\mathrm{\Psi}_{\mathbf{p}}, \ J_2 = \int d\mathbf{r}\mathrm{\Psi}^{*}_{\mathbf{p}'}e^{-i\mathbf{qr}}\mathrm{\Psi}_{\mathbf{p}}.
\end{equation}

In a case of a quadratic dispersion law, with regard to only long-wavelength lattice vibrations, one can take the electron wave function in a form of plane wave
\begin{equation}
   \label{eq:eq060}
  \psi_{\mathbf{p}}\left(\mathbf{r}\right)=\dfrac{1}{\sqrt{L^2}}e^{\frac{i\mathbf{pr}}{\hbar}}.
\end{equation}

Matrix element (\ref{eq:eq58}), taken on this function, is well-known and it is presented in, for example, in \cite{36Bonch}. In material, which is described by Rashba Hamiltonian, the electron wave function is given by the next expression (in long-wavelength approximation too) \cite{36Bercioux}:
\begin{equation}
   \label{eq:eq60}
 \mathrm{\Psi}_{\pm}=\dfrac{1}{\sqrt{2 L^2}}\begin{pmatrix} 1 \\ \pm ie^{i\theta} \end{pmatrix} e^{\frac{i}{\hbar} \left(p_x x+p_y y\right)},
\end{equation}
$\theta=\arctan\left(\dfrac{p_y}{p_x}\right)$  is an angle between the components of quasi-momentum. We chose a top subband, so the wave function becomes the next:
\begin{equation}
   \label{eq:eq61}
 \mathrm{\Psi}_{\mathbf{p}}(\mathbf{r})=\dfrac{1}{\sqrt{2}}\begin{pmatrix} 1 \\ ie^{i\theta} \end{pmatrix}\psi_{\mathbf{p}}(\mathbf{r}).
\end{equation}
Consequently, the expression of a scattering probability should differ from the expression, depicted in \cite{36Bonch}, only by values of integrals (\ref{eq:eq59}). After simple transformations one can obtain:
\begin{equation}
   \label{eq:eq62}
 \left\langle \mathbf{p}',n'\left|H'\right|\mathbf{p}, n \right\rangle = \dfrac{1+e^{-i(\theta'-\theta)}}{2} \left\langle \mathbf{p}',n'\left|H'\right|\mathbf{p}, n \right\rangle _0,
\end{equation}
where $\left\langle \mathbf{p}',n'\left|H'\right|\mathbf{p}, n \right\rangle _0$ - is a matrix element, taken on plane waves ({\ref{eq:eq060}}). So a square of the matrix element, immediately incoming into (\ref{eq:eq55}), is
\begin{equation}
   \label{eq:eq63}
 \left|\left\langle \mathbf{p}',n'\left|H'\right|\mathbf{p}, n \right\rangle\right|^2 = \dfrac{1+\cos\phi}{2}\left|\left\langle \mathbf{p}',n'\left|H'\right|\mathbf{p}, n \right\rangle _0\right|^2.
\end{equation}
In our subsequent investigations (during the calculation of probability of electron transition from the state $\mathbf{p}$ to every possible states with quasi-momentum $\mathbf{p}'$) we shall assume $Ox$ axis directed along the vector $\mathbf{p}$, so angle difference $\theta'-\theta$ should be equal to polar angle $\phi$: $\theta'-\theta = \phi$. Thus, formulae of scattering probability in a case of calculation of matrix elements on eigenfunctions of Rashba Hamiltonian should differ from that, calculated on plane waves, only by coefficient $\dfrac{1+\cos\phi}{2}$ (taking into account, of course, that we consider the two-dimensional problem). General expression for charge carriers probability takes a form:
\begin{align}
   \label{eq:eq64}
  &W(\mathbf{p},\mathbf{p}')=\dfrac{1+\cos\phi}{2}\cdot\dfrac{2\pi}{\hbar}\cdot\left|C_{\mathbf{q}}\right|^2\cdot \nonumber \\
  \nonumber \\
  &\cdot \left(n_\mathbf{q}\delta\left(\varepsilon_{\mathbf{p}'}-\varepsilon_{\mathbf{p}}-\hbar\omega_\mathbf{q}\right)
  \delta_{\mathbf{q},(\mathbf{p}'-\mathbf{p})/\hbar} + \right. \nonumber \\
  \nonumber \\
  &\left.+\left(n_\mathbf{q}+1\right)\delta\left(\varepsilon_{\mathbf{p}'}-\varepsilon_{\mathbf{p}}+\hbar\omega_\mathbf{q}\right)
  \delta_{\mathbf{q},(\mathbf{p}-\mathbf{p}')/\hbar}\right). 
\end{align}
In this expression $\left|C_{\mathbf{q}}\right|^2=\dfrac{\hbar B^2(\mathbf{q})}{2L^2\rho\omega_{\mathbf{q}}}$ is a constant of electron-phonon interaction, $\omega_{\mathbf{q}}$ is a cyclic frequency of phonon, $\rho$ is a surface density of material, coefficient $B(\mathbf{q})$ corresponds to a potential of deformation, it depends on type of phonons (we taking in account only nonpolar acoustical and optical phonons), $n_{\mathbf{q}}$ is a phonon distribution function, $\varepsilon(\mathbf{p})$ is an energy of electron (in our case it described by (\ref{eq:eq2})). Because of we are interesting in motion of electrons under the influence of comparatively weak electric fields and temperatures, we shall assume phonon distribution to be close to equilibrium, so $n_{\mathbf{q}}$ can be described by Plank function: $n_{\mathbf{q}}=n_q=\dfrac{1}{\exp\left((\hbar\omega_q)/(kT)\right)-1}$. Note, that similar to (\ref{eq:eq64}) expression of charge carriers probability on nonpolar phonons in two-dimensional material with linear on quasi-momentum term in energy spectrum was derived in \cite{36Vasko} (for the case of graphene).

\subsection{Scattering probability in a unit of time of charge carriers on acoustical phonons}
\label{sec:4.2}

At temperatures less or equal to $100K$ an acoustical phonon energy is much less than representative energies of electron in initial and finishing states and energy of thermal motion: $\hbar\omega_\mathbf{q}\ll\varepsilon_{\mathbf{p}}, \varepsilon_{\mathbf{p}'}, kT$. So in arguments of Dirac delta-function in (\ref{eq:eq64}) we can neglect a phonon energy as well as we can reduce phonon distribution function to a form $n_q=\dfrac{kT}{\hbar\omega_q}$. In a case of charge carriers scattering on long-wavelength nonpolar acoustical phonons a coefficient $B^2(\mathbf{q})=D_a^2q^2$, where $D_a$ is a constant of deformation potential which typical value is about several eV. Since in long-wavelength approximation we deal with longitudinal lattice oscillations, we can represent cyclic frequency of acoustical phonons in a form $\omega_q=sq$, where $s$ is a speed of sound. In a such a way constant of electron-phonon interaction is $\left|C_\mathbf{q}\right|^2=\dfrac{\hbar D_a^2 q}{2L^2\rho s}$. Consider a unit as negligible compared to $n_q$ in the second term of (\ref{eq:eq64}), we obtain an expression for the probability of electron scattering by acoustic phonons in the following form:
\begin{align}
   \label{eq:eq65}
   &W_{ac}(\mathbf {p,p'})=W_{ac}(\mathbf {p',p})=\nonumber \\
   \nonumber \\
   &=\dfrac{1+\cos\phi}{2}\cdot\dfrac{2\pi D_a^2 kT}{\hbar L^2\rho s^2} \cdot \delta(\varepsilon(\mathbf {p'})-\varepsilon(\mathbf{p})).
\end{align}

To find a total probability $W(\mathbf{p})$ of electron scattering by phonons with the momentum of $\mathbf{p}$ on one of the acoustic branches one should integrate (\ref{eq:eq65}) over the final momentum $\mathbf{p'}$, as well as the configuration space:
\begin{equation}
   \label{eq:eq66}
   W(\mathbf{p})=\dfrac{L^2}{(2\pi\hbar)^2}\int\it W(\mathbf {p, p'})d\mathbf{p'}.
\end{equation} 
Going over in a polar coordinate system in momentum space and substituting the explicit expression of the energy spectrum (\ref{eq:eq2}), we obtain the following expression for the total probability:
\begin{align}
\label{eq:eq67}
   &W_{ac}(\mathbf{p}) = \dfrac{2\pi D_a^2 kT}{\hbar L^2\rho s^2}\cdot  \dfrac {L^2}{(2\pi\hbar)^2}  \int\limits_{0}^{2\pi} d\phi\dfrac{1+\cos\phi}{2} \nonumber \\
  &\cdot\int\limits_{0}^{\infty}dp'~{p'}~\delta\left({\dfrac{{p'}^2}{2m}+\dfrac{\alpha}{\hbar}p'-\dfrac{p^2}{2m}-\dfrac{\alpha}{\hbar}p}\right).
\end{align}
Using the rule (see, for example, \cite{37Arfken})
\begin{equation}
   \label{eq:eq68}
   \delta\left[f(x)\right]=\dfrac{1}{|\partial f/\partial x|}\delta(x-x_0),
\end{equation}
where $x_0$ is defined by condition $f(x_0)=0$, one obtains
\begin{equation}
   \label{eq:eq69}
   \delta\left({\dfrac{{p'}^2}{2m}+\dfrac{\alpha}{\hbar}p'-\dfrac{p^2}{2m}-\dfrac{\alpha}{\hbar}p}
   \right)=\dfrac{\delta (p'-p)}{\dfrac{p'}{m}+\dfrac{\alpha}{\hbar}} 
\end{equation}
The final expression for the total probability of electron scattering by acoustic phonons becomes a form:
\begin{equation}
   \label{eq:eq70}
   W(p)=\dfrac {D_a^2 kT }{2\rho s^2\hbar^3}\cdot\dfrac{p}{\dfrac{p}{m}+\dfrac{\alpha}{\hbar}}
\end{equation}
The scattering of electrons by acoustic phonons is assumed to be elastic, i. e. the absolute value of the momentum is unchanged. During the modelling the recalculation of projections of momentum is performed, for which purpose the values of an angle $\phi$ are raffled off. It is convenient to introduce dimensionless variables: $\mathbf{p}\rightarrow \mathbf{p}\sqrt{2mkT}$, which we use already in Section \ref{sec:2}. The probability of charge carrier scattering by acoustic phonons can be presented as
\begin{equation}
   \label{eq:eq71}
   W_{ac}(p)=A_{ac}\cdot\dfrac{p}{p+X}, \ A_{ac}=\dfrac{D_a^2kTm}{2\rho s^2\hbar^3}.
\end{equation}

\subsection{Scattering probability in a unit of time of charge carriers on optical phonons}
\label{sec:4.3}

Let us to write down the expression for an electron transition probability $W(\mathbf {p,p'})$ from the state with momentum $\mathbf{p}$ to the state with momentum $\mathbf{p'}$ under the influence of the scattering on optical lattice oscillations. For simplicity we shall assume phonons to be dispersionless with energy $\hbar\omega_o$. Consider the case when the condition $\dfrac{\hbar\omega_o}{\it kT}\gg 1$  is satisfied. For many materials this condition holds well at temperature about 100 K \cite{36Bonch}. In this case one obtains $n_q\ll 1$, and only the second term in (\ref{eq:eq64}), which describes the scattering of electron with the phonon emission, is substantial. In approximation of a deformation potential for optical lattice vibrations the coefficient $B^2(\mathbf{q})=D_o^2$, $D_o\sim 10^9 \mathrm{eV/cm}$, so a constant of electron-phonon interaction is ${|C_\mathbf{q}|}^2=\dfrac{\hbar D_o^2}{2L^2\rho\omega_o}$. Thus, the expression for the probability $W(\mathbf {p, p'})$ takes a form:
\begin{equation}
   \label{eq:eq72}
   W_{opt}(\mathbf {p, p'})=\dfrac{1+\cos\phi}{2}\cdot\dfrac{\pi D_o^2}{L^2\rho\omega_o}\cdot\delta\left(\varepsilon(\mathbf {p'})-\varepsilon(\mathbf {p})+\hbar\omega_o\right).
\end{equation}
As in the case of carrier scattering by acoustic phonons, in order to calculate the total scattering probability $W(\bf p)$ of electron with the momentum $\bf p$ on the optical lattice vibrations we integrate an expression (\ref{eq:eq72}) on all final momentum values $\mathbf{p}'$ and on configuration space.
\begin{align}
\label{eq:eq73}
   &W_{opt}(\mathbf{p}) = \dfrac{\pi D_o^2}{L^2\rho\omega_o}\cdot \dfrac {L^2}{(2\pi\hbar)^2}\cdot \int\limits_{0}^{2\pi}d\phi\dfrac{1+\cos\phi}{2}\cdot \nonumber \\  
   &\cdot\int\limits_{0}^{\infty} dp'~p'~\delta\left(\dfrac{{p'}^2}{2m}+\dfrac{\alpha}{\hbar}p'-\dfrac{p^2}{2m}-\dfrac{\alpha}{\hbar}p+\hbar\omega_o \right). 
\end{align}
Using the rule (\ref{eq:eq68}), one transforms (\ref{eq:eq73}):
\begin{equation}
   \label{eq:eq74}
   W_{opt}(p)=\dfrac{D_o^2}{4\rho\omega_o\hbar^2}\cdot
   \int\limits_{0}^{\infty}\dfrac{dp'~p'}{\dfrac{p'}{m}+\dfrac{\alpha}{\hbar}}\delta\left(p'-p_0'\right),
\end{equation} 
where $p_0'$ is a solution of equation
\begin{equation}
   \label{eq:eq75}
  \dfrac{{p'}^2}{2m}+\dfrac{\alpha}{\hbar}p'-\dfrac{p^2}{2m}-\dfrac{\alpha}{\hbar}p+\hbar\omega_o=0,
\end{equation}
\begin{equation}
   \label{eq:eq76}
   p_0'=-\dfrac{m\alpha}{\hbar}\pm\sqrt{\left(p+\dfrac{m\alpha}{\hbar}\right)^2-2 m\hbar\omega_o}
\end{equation}
The value $p_0'$ must be non-negative real number, so one should choose $p_0'=-\dfrac{m\alpha}{\hbar}+\sqrt{\left(p+\dfrac{m\alpha}{\hbar}\right)^2-2 m\hbar\omega_o}$. The condition $p_0'\geq 0$ can be converted to a form
\begin{equation}
   \label{eq:eq77}
   \dfrac{p^2}{2m}+\dfrac{\alpha}{\hbar} p \geq\hbar\omega_o.
\end{equation}
The expression (\ref{eq:eq77}) has a simple physical meaning: the scattering of an electron with emission of phonon, which is taken into account in (\ref{eq:eq72}), is possible if the energy of the electron is not less than the phonon energy. Passing to the non-dimensional quasi-momentum in the same way as it was done in the case of scattering by acoustic vibrations $(p\rightarrow p\sqrt{2mkT})$, we obtain an expression for the scattering probability of electron with momentum $\mathbf{p}$ on optical phonons in the following form:
\begin{align}
\label{eq:eq78}
    &W_{opt}(\mathbf {p})=\nonumber \\ 
    &=\left\{\begin{array}{rcrc}
         &0,  &p'_0<0;\\
         &A_{opt}\left(1-\dfrac{X}{\sqrt{\left(p+X\right)^2-\beta}}\right), &p'_0\geq 0.\end{array}\right.
\end{align}
Here 
\begin{equation}
   \label{eq:eq79}
   p'_0=\sqrt{(p+X)^2-\beta}-X
\end{equation}
is an absolute value of quasi-momentum of electron after scattering by optical phonon, $A_{opt}=\dfrac{D_o^2 m}{4\rho\omega_o\hbar^2}$, $\beta =\dfrac{\hbar\omega_o}{kT}$, $X=\dfrac{m\alpha}{\hbar\sqrt{2mkT}}.$

\section{Modelling of a distribution of charge carriers on quasi-momentum value and polar angle}
\label{sec:5}

For a correct use of statistical methods one needs to specify the initial distribution of quasi-momenta. As in a Sect. {\ref{sec:2}}, we shall assume electron gas is non-degenerated so its distribution function has a Boltzmann form (\ref{eq:eq9}). A probability that electron is localized in an element of quasi-momentum space $dp_xdp_y$ can be presented in a form $dw=f\left(p_x,p_y\right)dp_xdp_y$. Passing into a polar coordinate system one obtains
\begin{equation}
\label{eq:eq80}
dw=A_{\mathrm{norm}}\exp\left(-\dfrac{p^2}{2mkT}-\dfrac{\alpha p}{\hbar kT}\right)pdpd\phi,
\end{equation}
where $A_{\mathrm{norm}}$ is described by (\ref{eq:eq10}). Going to dimensionless quantities $p\rightarrow p/\sqrt{2mkT}$ and having regard to a factor $1/(2\pi\hbar)^2$ in summation on quasi-momentum space, after integration on polar angle one obtains
\begin{equation}
\label{eq:eq81}
dw=\dfrac{2\exp\left(-p^2-2Xp\right)p}{1-X\sqrt{\pi}e^{X^2}(1-\textrm{erf(X)}}dp.
\end{equation}
So a probability density of distribution on quasi-momentum absolute value
\begin{equation}
\label{eq:eq82}
f(p)=\dfrac{2\exp\left(-p^2-2Xp\right)p}{1-X\sqrt{\pi}e^{X^2}(1-\mathrm{erf(X)}}.
\end{equation}
To obtain a random variable having a distribution function (\ref{eq:eq82}), we use the von Neumann's method (see, for example, \cite{38Kashurnikov}). The expression (\ref{eq:eq82}) possesses a maximum value at
\begin{equation}
\label{eq:eq83}
p_0=\dfrac{-X+\sqrt{X^2+2}}{2},
\end{equation}
and this value is
\begin{equation}
\label{eq:eq84}
f_{\mathrm{max}}=\dfrac{-X+\sqrt{X^2+2}}{1-X\sqrt{\pi}e^{X^2}(1-\mathrm{erf}(X))}.
\end{equation}
To get a needed distribution of $p$ one should carry out the next procedure. Two random numbers are raffled out: a first number $p$ is a quasi-momentum absolute value, it should be a uniformly distributed random number from an interval $[ 0, p_{\mathrm{max}} ]$, and a second number $f_r$ is a uniformly distributed random value from an interval $[0,f_{\mathrm{max}}]$. If $f_r<f(p)$, we have found needed value $p$, else we should raffle out quantities $p$ and $f_r$ once again. Values $p$ derived by this means should fill an area under the graphic of $f(p)$ and depict an initial distribution of a quasi-momentum absolute value.

During the modelling we should each time after collision define a quasi-momentum absolute value and angle between quasi-momentum vector before and after collision. Expression (\ref{eq:eq65}), (\ref{eq:eq72}) describe probability of scattering of electron on acoustical and optical phonons, respectively. Last terms in these expressions are the laws of conservation of energy which are used for determination of an absolute value of the electron quasi-momentum after scattering (in a case of acoustical phonons this value is assumed to be equal to initial quasi-momentum of electron, in a case of optical phonons this value is defined by (\ref{eq:eq79})). Expressions (\ref{eq:eq65}), (\ref{eq:eq72}) are derived in approximation of long-wavelength phonons so phonon wave vector $\mathbf{q}$ should satisfy the condition
\begin{equation}
\label{eq:eq85}
\dfrac{2\pi}{q}\gg a,
\end{equation}
where $a\sim 10^{-8}\mathrm{cm}$ is a lattice spacing. Absolute values of momentum of electron in initial and final states and momentum of phonon are the quantities of the same order. The condition (\ref{eq:eq85}) for dimensionless quasi-momentum takes a form 
\begin{equation}
\label{eq:eq86}
p\ll \dfrac{2\pi\hbar}{a\sqrt{2mkT}}.
\end{equation}
At temperatures $T\simeq70K$, electron mass in order of free electron mass this inequality is well satisfied throughout the region of interest of momenta.

A polar angle, which corresponds to the electron after collision, can be found based upon the fact that   
first terms in (\ref{eq:eq65}), (\ref{eq:eq72}) can be treated as a density of probability of electron scattering by polar angle $\phi$:
\begin{equation}
\label{eq:eq87}
g(\phi)=\dfrac{1+\cos\phi}{2}.
\end{equation}
So to define an angle between the quasi-momentum vectors we use von Neumann's method: we raffle out two quantities: $\phi$ which is uniformly distributed in the interval $(0, \ \pi)$, and $g_r$, uniformly distributed in the interval $(0, \ 1)$. If $g_r<g(\phi)$, we have found desired value $\phi$, else we should raffle out these values again.

\section{Numerical results}
\label{sec:6}

\begin{figure}
\includegraphics[scale=0.45]{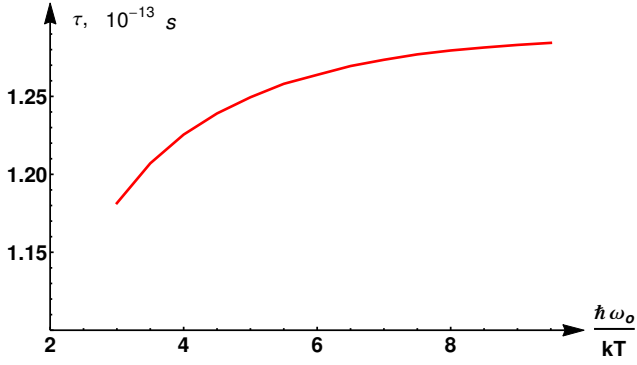}
\caption{Dependence of a mean relaxation time on an energy of optical phonons ($E_x=E_y=E_{yc}=3.0$)}
\label{fig:5}
\end{figure}

First of all, it is interesting to calculate the mean free time of charge carriers and find its dependence on the electric field, type and energy of phonons. A presence of only acoustical phonons in system can't lead to an appearance of transverse direct current, which confirms the results of \cite{6Tulkina}, obtained for the same problem in graphene. About the absence of the effect we conclude by the violent increasing of a direct current density variations at small values of optical phonon energy. The effect don't occur at $\alpha=0$, which is physically reasonable, because for a transverse direct current appearance an energy spectrum non-additivity is needed. On Figure \ref{fig:5} there is presented a dependence of a mean relaxation time $\tau$ on an energy of optical phonons (we use non-dimensional quantity $\beta =\hbar\omega_o/kT$). One can see that mean relaxation time is about $\tau\simeq 10^{-13}\mathrm{s}$ and weakly increase with phonon energy rise. The expression of probability of charge carriers scattering (\ref{eq:eq78}) is derived in condition of $\beta\gg1 \ $, so we should use as much as possible values of the ratio of the phonon energy to temperature. The value of mean relaxation time $\tau=10^{-13}\mathrm{s}$, needed to effect appearance, is stringent condition for two-dimensional materials, so we shall use in further investigations $\beta=8.2$, which at $T=70 \mathrm{K}$ is equivalent to the optical phonon energy $\hbar\omega_o\simeq 0.05 \mathrm{eV}$.

\begin{figure}
\includegraphics[scale=0.45]{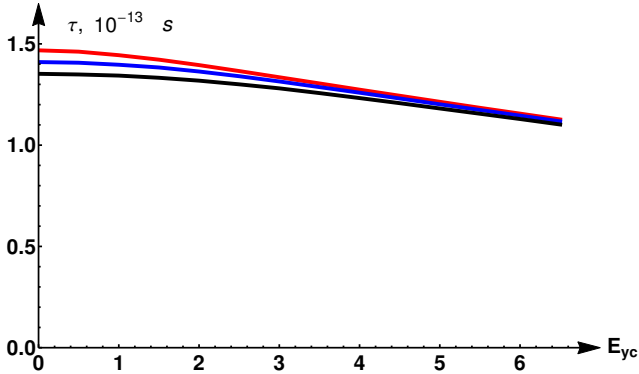}
\caption{Dependence of a mean relaxation time on the strength of constant components of electric field. Red curve correspond to $E_x=E_y=1.0$, blue curve correspond to $E_x=E_y=2.0$, black curve correspond to $E_x=E_y=3.0$}
\label{fig:6}
\end{figure}

On Figure \ref{fig:6} there are the dependencies of the mean relaxation time $\tau$ on the constant component of electric field $E_{yc}$. One can see that $\tau$ very weakly depends on the components of electric field strength of elliptically polarized wave $E_x$ and $E_y$. With a rise of a constant component of electric field $E_{yc}$ the mean relaxation time slowly decrease. Graphics on Fig. {\ref{fig:5}} and {\ref{fig:6}} show that approach on constant collision frequency $\nu=1/\tau$, which is used in Section \ref{sec:2}, is a good approximation for investigated problem.

\begin{figure}
\includegraphics[scale=0.45]{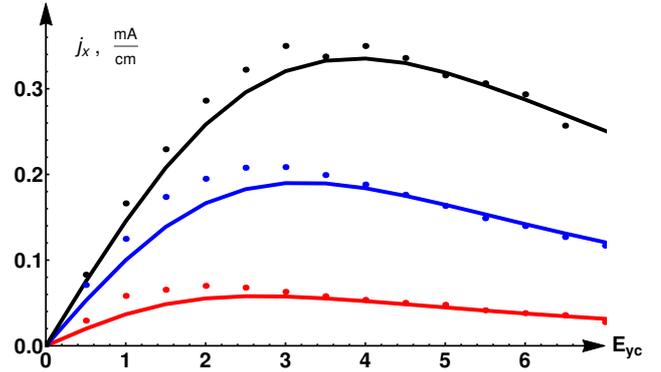}
\caption{Dependence of a direct component of transverse current density $j_x$ on the strength of constant component of electric field $E_{yc}$. Red curve and points correspond to $E_x=E_y=1.0$, blue curve and points correspond to $E_x=E_y=2.0$, black curve and points correspond to $E_x=E_y=3.0$}
\label{fig:7}
\end{figure}

On a Figure \ref{fig:7} the dependence of the transverse direct current density $j_x$ on $E_{yc}$ at different values of $E_x$ and $E_y$ is presented. Solid lines on the figure correspond to the calculations on the base of the constant collision frequency approximation, points correspond to results of Monte Carlo simulations. All graphics, respective to the constant collision frequency approximation, are plotted at $\eta=\dfrac{\nu}{\omega}=\dfrac{1}{\tau\omega}=10$, so the best coincidence of results, obtained by different methods, one can see at $E_{yc}\geq3$. At these values of $E_{yc}$, according to Figure {\ref{fig:5}}, mean collision frequency nears to $\nu=1.0\cdot 10^{13} s^{-1}$.

\begin{figure}
\includegraphics[scale=0.45]{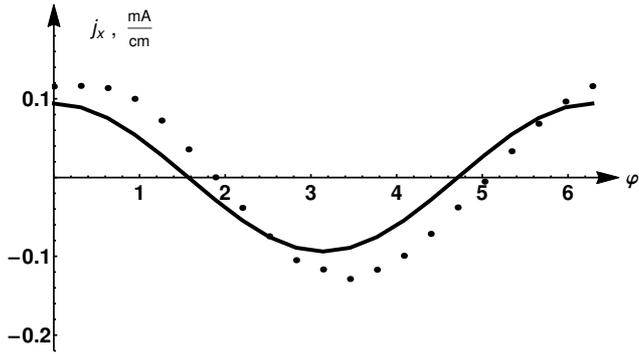}
\caption{Dependence of a direct component of transverse current density $j_x$ on the phase shift $\varphi$ between the components of elliptically polarized wave. Graphics correspond to $E_x=E_y=E_{yc}=1.5$}
\label{fig:8}
\end{figure}

The dependence of a transverse direct current density $j_x$ on phase shift $\varphi$ between components of the elliptically polarized wave is presented on a Figure \ref{fig:8}. Solid lines correspond to the approach of a constant collision frequency $\nu=10^{13} \mathrm{s^{-1}}$, points -- to the Monte Carlo simulations. One can see, that in approach of a constant collision frequency with a high degree of accuracy $j_x\sim\cos(\varphi)$. Same phase shift dependence is a feature of the effect of the direct current generation in transverse direction to a vector of a constant component of electric field strength under the influence of elliptically polarized electromagnetic wave which is incident normally to the surface of the sample of material with non-additive spectrum. This result coincide with them obtained in {\cite{5Marchuk}, \cite{8GSLKontchenkov}} for the case of graphene and graphene superlattice, respectively, using the same approximation. Monte Carlo simulations gives to another type of dependence: $j_x\sim\cos\left(\varphi+\Theta\left(E_x,E_y,E_{yc},\omega_0\right)\right)$. Analogous result was obtain for the case of graphene in {\cite{6Tulkina}} where there is a macroscopic explanation of this phase shift dependence.

The modelling was carried out at next values of parameters: a constant of deformation potential of acoustic phonons is $D_a=5~ \mathrm{eV}$, a constant of deformation potential of optical phonons is $D_o=2\cdot 10^9 ~\mathrm{eV}/\mathrm{cm}$, a mass of charge carriers is equal to the mass of free electron, a surface density of the sample is $\rho=3\cdot 10^{-6}~\mathrm{g/cm^2}$, a speed of sound is $s=10^5 ~\mathrm{cm/s}$, a surface concentration of charge carriers is $n=10^{10} ~\mathrm{cm^{-2}}$, a temperature is $T=70 \mathrm{K}$. An energy of an optical phonon in simulations, results of which is presented on Figures {\ref{fig:6}}, {\ref{fig:7}} and {\ref{fig:7}}, is $\hbar\omega_o=0.05 ~\mathrm{eV}$. A cyclic frequency of elliptically polarized wave is $\omega=10^{12} ~s^{-1}$. At these values coefficients $A_{ac}\approx A_{opt}=9.5\cdot 10^{12} ~s^{-1}$, so differences in frequency of a charge carriers scattering on acoustical and optical phonons are defined only owing to the presence of minimal energy, required an electron to emit the optical phonons. As shown by simulation, scattering of the electron on an acoustical phonon takes place five times more often then scattering on optical phonon. Modelling was carried out on time interval with length of 2000 in units $1/\omega$, time step was $2\cdot 10^{-4}$, minimal number of particles 1000 (in cases when we get great dispersion, we repeat our calculations several times and average out values of direct current density on these iterations). A typical number of collisions on acoustical phonons was about 14000. Motion equations {\ref{eq:eq6}} are solved by the Runge-Kutta method of the fourth order.

Initial quasi-momentum distribution $\mathbf p_0$ is chosen by mean, presented in Section {\ref{sec:5}}. Our calculations show that initial distribution has a significant impact only for collisionless kinetic processes and in a case of investigations of different types of scattering it exerts a weak influence on the modelling result. In our modelling we use parallel calculation on GPU (graphics processing unit) on the base of NVidia CUDA technology. We have realized a XORSHIFT random generator \cite{39Xorshift} in kernel function. Each generator is initialized accordingly of thread and block indexes of kernel. Using own random generator in each kernel allows us to perform all calculations for each particle (which is specified by initial conditions) on GPU. We find an average electron velocity on ensemble as an arithmetic mean of the velocity values in each time moment, calculated by different kernels. Then we average electron velocity on time. For work with the GPU we used Python and its package PyCUDA {\cite{40PyCUDA}}.

Using the Monte Carlo modelling allows us to calculate a mean relaxation time, to find its dependencies on electric field components and on an energy of optical phonon. Using Monte Carlo simulations we could directly verify the results of the approach of a constant collision frequency. These calculations allow us to define more exactly the dependence of direct current on phase shift between the components of elliptically polarized wave.

In our investigations we examine values of electric field strengths order of tens of SGSE units, which is, strictly speaking, great values. The reason of this choice is large value of dispersion of direct current in weak electric fields, if we use Monte Carlo method for these calculations. Method of a constant collision frequency, used in Section {\ref{sec:2}}, gives good results in a case of weak fields. Furthermore, this method can be used in case when $\nu/\omega\leq 1$.

\section{Conclusions}
\label{sec:7}

On the base of approach of a constant collision frequency and using semiclassical Monte Carlo simulations the effect of the appearance of transverse direct current in material with a Rashba Hamiltonian under the influence of an elliptically polarized wave and a constant electric field is studied. Modelling on the base of Monte Carlo method allows us to define a mean relaxation time and its dependence on components of the electric field and on the energy of optical phonons. This modelling shows that the conditions of application of a constant relaxation time approximation are satisfactorily hold true in the considered problem. A comparison of direct current component calculations on the base of these two approach is carried out.



%
%

\begin{acknowledgements}
The work is supported by the grant of RFBR No. 13-02-97033 r\_povolzh'e\_a, and it is carried out under the financial support of the Ministry of Education and Science of the Russian Federation for perform public works in the field of scientific activities of the base portion of the state task number 2014/411 (Project Code: 522). The work is performed using the NVIDIA Kepler coprocessor, acquired within the framework of The Program of a Strategical Development of the Volgograd State Technical University. 
\end{acknowledgements}



\end{document}